\documentclass[
aps,%
10pt,%
final,%
notitlepage,%
oneside,%
twocolumn,%
nobibnotes,%
nofootinbib,%
superscriptaddress,%
noshowpacs,%
centertags]%
{revtex4}

\begin{document}

\selectlanguage{english}

\title{Anisotropy of the Space Orientation of Radio Sources. II:
The Axis Distribution Function }
\author{\firstname{V.~R.}~\surname{Amirkhanyan}}

\affiliation{Sternberg Astronomical Institute, Universitetskii pr.
13, Moscow, 119992 Russia}

\received{March 23, 2009}%
\revised{May 20, 2009}%

\begin{abstract}
An analysis of the position angles distribution of 10461 extended
radio sources shows that the spatial orientation of the axes of
these objects is anisotropic: they avoid the direction towards the
Celestial Pole and are mostly oriented in the equatorial
direction. The ratio of the probability densities of the
orientation in these two directions is  0.68. The probability that
the sky distribution of axes is isotropic is less than 0.00004.
This conclusion is consistent with the results of the analyses of
spatial distribution of galactic normals orientation in the
galaxies from the UGC, ESO, and FGC catalogs.

\end{abstract}
\pacs{98.65.Dx, 98.54.Aj, 98.54.Gr, 98.62.Ve}

\maketitle
\section{INTRODUCTION}

Reinhardt was the first to study the spatial orientation of
galaxies over the entire sky \cite{1:Amirkhanyan2_n_en}. He used
the catalog of galaxies compiled by
Brawn~\cite{2:Amirkhanyan2_n_en}, a pioneer of such
investigations. Reinhardt showed that the position angles
distribution of the major axes of galaxies is not equiprobable and
that the tilts of galaxies with respect to the observer's line of
sight are inconsistent with the hypothesis of equiprobable
distribution of galaxy normals orientation. The next, very
important step, was made by Peter Nielson, who not only compiled
the extensive Uppsala catalog of galaxies of the Northern
Hemisphere ~\cite{3:Amirkhanyan2_n_en}, but also showed with high
statistical significance that the distribution of position angles
of galaxies is not equiprobable~\cite{4:Amirkhanyan2_n_en}.
Lauberts extended the search for galaxies to the Southern
Hemisphere and compiled the  ESO
catalog~\cite{5:Amirkhanyan2_n_en}. He analyzed the orientation of
UGC and ESO objects to construct the probability density function
of galaxy normals spatial orientation in form of a
three-dimensional ellipsoid. Two axes of the ellipsoid with the
lengths of 1 and 0.933 lie practically in the equatorial plane.
The third, shortest, axis (0.778) points towards the Celestial
Pole. Mandzhos et al.~\cite{6:Amirkhanyan2_n_en}, too, analyzed
the orientation of UGC and ESO galaxies and showed that the
cylindrical model of spatial orientation agrees best with
observations. The axis of the cylinder, perpendicularly to the
lateral surface of which the galactic normals are preferentially
oriented, points towards the Celestial Pole. Parnovsky et
al.~\cite{7:Amirkhanyan2_n_en}, apart from the UGC and ESO
objects, included into the analysis their FGC catalog of edge-on
galaxies, and showed with high statistical significance that the
spatial orientation of galaxy normals  is anisotropic: there are
zones of avoidance (Celestial Pole) and zones of preferred
orientation (the equator). Thus the above authors demonstrated
convincingly that the spatial distribution of galaxy normals is
anisotropic. Note that normals concentrate towards the equator and
avoid the direction towards the Celestial Pole. Such a surprising
coincidence is alarming. It is worthwhile to test this result on
an independent sample of objects. That is why we compiled, based
on the FIRST survey, a catalog of extended radio
sources~\cite{9:Amirkhanyan2_n_en}, the orientations of which we
analyze in this paper.

\section{SPATIAL DISTRIBUTION OF AXES. THE EXPERIMENT}
\label{intro:Amirkhanyan2_n_en}

Extended radio sources have a uniform structure with a
well-defined axis. Modern radio telescopes produce bona fide
images of radio sources out to several tens of thousands of Mpc,
whereas galaxies can be properly imaged out only to about
\mbox{200 Mpc}; at greater distances, the errors of the geometric
parameters of galaxies inferred from their image increase
substantially. In [9, Fig. 7c] we constructed the position angle
histogram of the catalogue radio sources. It follows from this
histogram that the probability that the distribution of angles is
equiprobable is less than 10$^{-7}$. The evident conclusion is
that the spatial orientation of radio sources is anisotropic. The
inverse statement is, generally speaking, not true: a ``flat''
distribution of position angles with respect to the Celestial Pole
does not prove the absence of anisotropy of spatial orientation.
Our current knowledge is insufficient to determine the spatial
orientation of a particular radio source, and we can only record
the celestial projection of its structure. Hence we have at our
disposal the information about the coordinates of the radio
sources, and the position angles of their projections with respect
to the Celestial Pole. We need to reconstruct the one-dimensional
distribution of position angles into a two-dimensional
distribution of axes of the radio sources. Let us make the two
following obvious statements:

\begin{figure*}[tbp]
\setcaptionmargin{5mm} \onelinecaptionsfalse \captionstyle{normal}
\centerline{\includegraphics[width=10cm, bb=57 173 549 547,
clip]{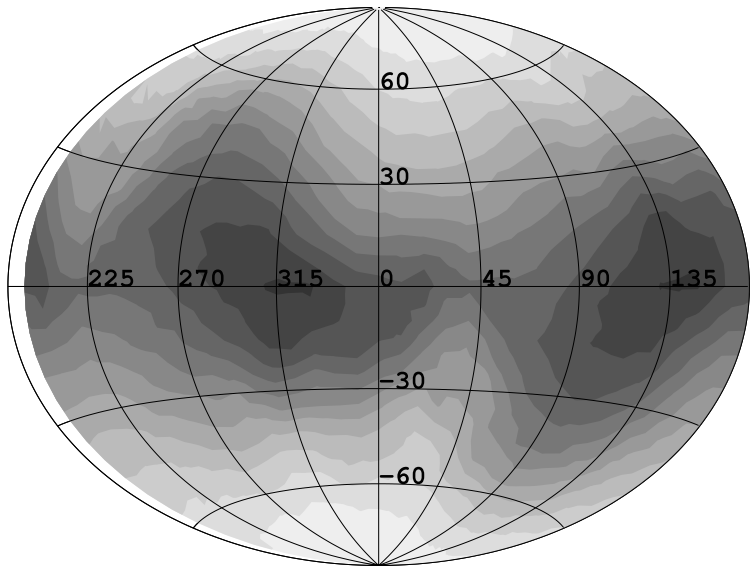}} \caption{Experimental map of the
non-uniformity of the radio sources' position angles distribution.
Transition from light to dark shades corresponds to the change of
the map level from the minimum towards the maximum.}
\label{mapex:Amirkhanyan2_n_en}
\end{figure*}

(1) if there is a celestial direction of  ``preferred
orientation'' of the radio sources, then the distribution function
of position angles with respect to this point should have the
maximum non-uniformity, as compared to the distributions of
position angles with respect to the neighboring points. Note that
the maximum and minimum of the function are close to the zero and
90$\degr$ position angles, respectively;

(2) if there is a direction in the sky, which the orientations of
radio sources ``tend to avoid'', then the distribution function of
position angles with respect to this point should have the maximum
non-uniformity and the maximum and minimum of the function should
be close to the 90$\degr$ and zero  position angles, respectively.

The more objects tend (or avoid) to orient themselves towards a
certain direction, the greater is the non-uniformity of the
histogram of position angles measured with respect to this
direction. These are the underlying ideas of the organization of
the algorithm used to construct the distribution function of the
spatial orientation of the axes. The \textbf{anisoreal} program
moves a point over the sky and, based on the ``knowledge'' of the
coordinates and position angles of the radio sources of the
catalog~\cite{9:Amirkhanyan2_n_en}, computes their position angles with respect to
the current point, draws the histogram of angles, and computes its
parameters. We characterize the non-uniformity of the histogram by
the amplitude S of its convolution with a single period of the
$\cos(2p)$ harmonics, where $p$ is the position angle counted in
the interval from 0 to 180 degrees. It is clear from the above
that the convolution reaches its positive and negative extrema at
the point of the preferred orientation and at the point of
avoidance, respectively. To pass to the amplitude of
non-uniformity, we normalize the convolution  \mbox{to}

$$
W=\sum_{i = 0}^{n-1} {\cos^2({2\pi{i}/n),} }\eqno(1)
$$

\noindent where $n$ is the number of bins in the histogram. The
program yields the map of the sky (in equatorial coordinates),
where the count at each point is equal to the non-uniformity
amplitude of the position angles histogram of all the catalog's
radio sources with respect to the given point (Fig.\,1). In this
map, the transition from lighter to darker shades corresponds to a
change of level from the minimum towards the maximum. Figures~2a
and 2b show the histograms at the extreme points of the map. The
negative extremum is located at (120$\degr$, 80$\degr$) and its
level is --53.4. The non-uniformity of the histogram at the point
of the positive extremum (140$\degr$, 0$\degr$) is equal to 46.2.
The dotted line shows the average level and the error bar computed
assuming that the distribution of position angles is equiprobable.
The probability that the histograms at the extreme points
represent a ``flat'' distribution is less than 10$^{-7}$. Figure~3
shows the cross section of the map along the 130$\degr$ meridian
(the solid line). The dotted line shows the cross section of
isotropic distribution. Its radius is equal to 581.17, i.e., the
average number of radio sources per histogram bin. The deviation
of isotropy is characterized by the amplitude of non-uniformity of
the histograms. In this figure, the y- and x-axes correspond to
the Polar axis and the equatorial plane, respectively. The map in
Fig.\,1 demonstrates that the axes of radio sources are
preferentially oriented towards the equatorial region and avoid
the direction towards the Celestial Pole. This surprising result
agrees with at least equally surprising results based on the
studies of UGC, ESO, and FGC galaxies. To check the operation of
the \textbf{anisoreal} program, we repeated the same computations
for objects of the Flat Galaxies Catalogue, that analyzed in
earlier paper~\cite{7:Amirkhanyan2_n_en}. As a result, we obtained
a map of non-uniformity of the normals position angle distribution
of these objects (Fig.\,4). This map shows that galactic normals,
like the axes of radio sources, avoid the direction towards the
Celestial Pole and prefer the equatorial region, matching the
results of Parnovsky et al.~\cite{7:Amirkhanyan2_n_en}.

\begin{figure*}[tbp]
\setcaptionmargin{5mm} \onelinecaptionsfalse \captionstyle{normal}
\centerline{
\includegraphics[width=12cm, bb=122 185 438 713,clip]{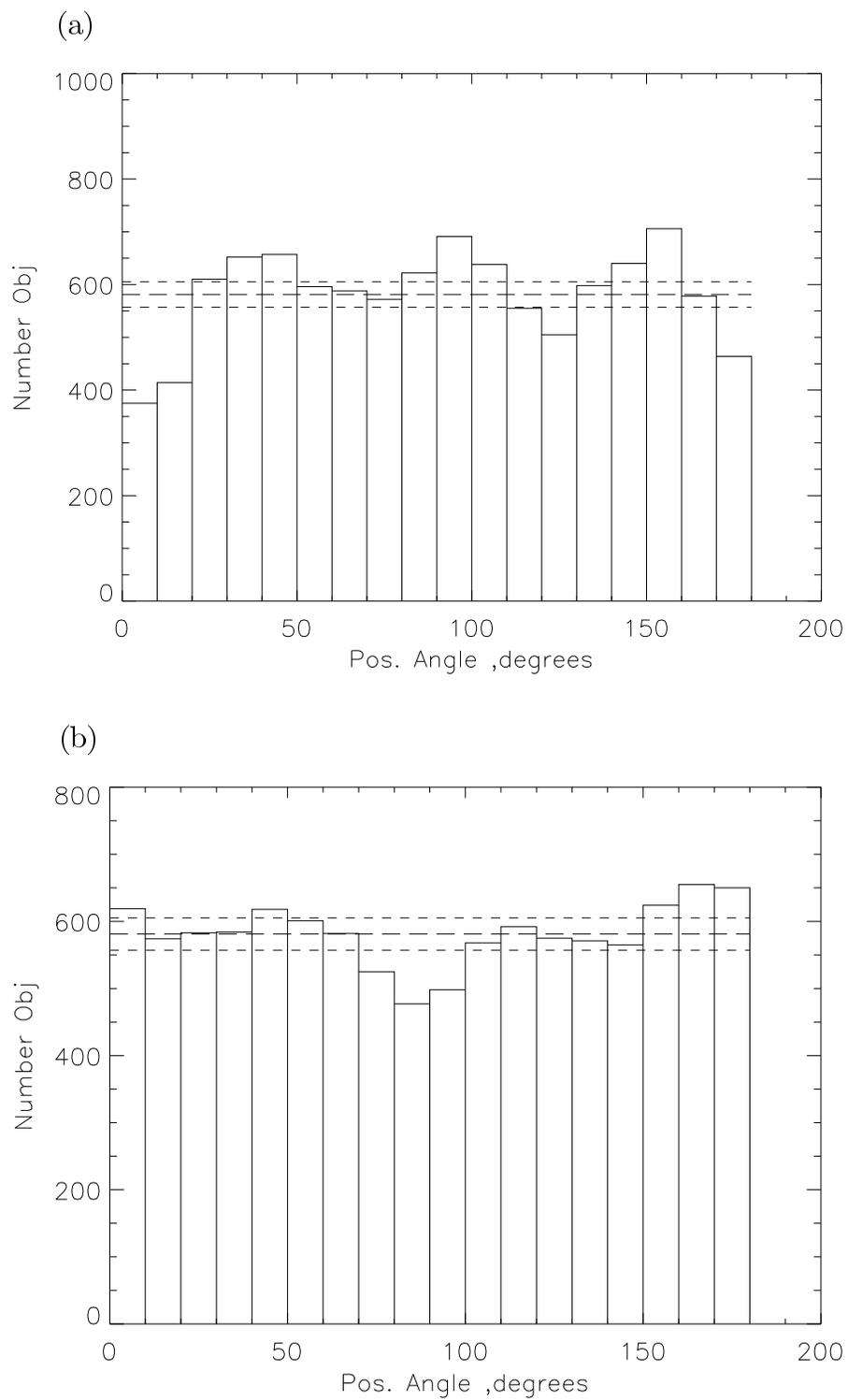}
}
\caption{(a) The distribution of position angles with respect to
the point of the negative extremum (120$\degr$, 80$\degr$); the
non-uniformity of the histogram is 53.4. (b) The distribution of
position angles with respect to the point of the positive extremum
(140$\degr$, 0$\degr$); the non-uniformity of the histogram is
46.2.} \label{gisti:Amirkhanyan2_n_en}
\end{figure*}

\begin{figure*}[tbp]
\setcaptionmargin{5mm} \onelinecaptionsfalse \captionstyle{normal}
\centerline{\includegraphics[width=10cm, bb=31 160 430 550,
clip]{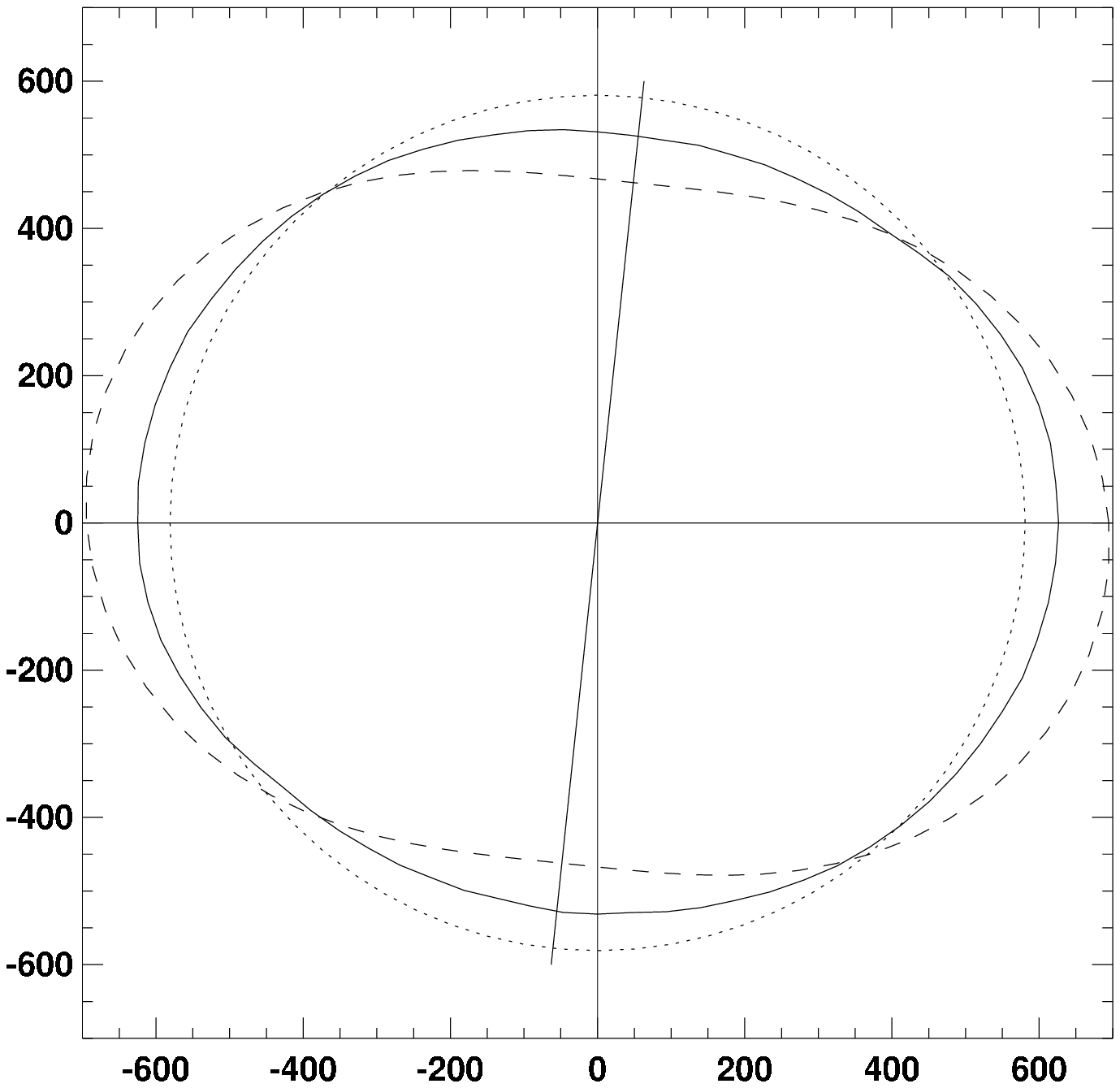}} \caption{The cross section of the
experimental map of non-uniformity (the solid line) and the
distribution function of the radio sources' axes orientation. The
dotted line corresponds to the isotropic distribution function for
${\rm A}=0$. The dashed line is the same function for ${\rm
A}=0.5$. The angle between the anisotropy axis and the Y axis is
equal to 6$\degr$.} \label{srez:Amirkhanyan2_n_en}
\end{figure*}

\section{SPATIAL DISTRIBUTION OF AXES. THE MODEL}

To associate the resulting map with real anisotropy, let us construct a model of anisotropy and learn how
to estimate its parameters. Here is the sequence of operations:

(1) we generate a list of the coordinates of  10461 objects
uniformly distributed over the sky. We can also use the
coordinates of the objects of the real catalog~\cite{9:Amirkhanyan2_n_en};

(2) we choose the following simple function of the spatial
distribution of the axes defined by the formula:

$$
P(\varphi,\omega)d{\varphi}d{\omega}=\frac{1+A\cos^2\omega}{\pi(1+A/2)}d{\varphi}d{\omega}. \eqno(2)
$$

\begin{figure*}[tbp]
\setcaptionmargin{5mm} \onelinecaptionsfalse \captionstyle{normal}
\centerline{\includegraphics[width=10cm, bb=58 176 552 549,
clip]{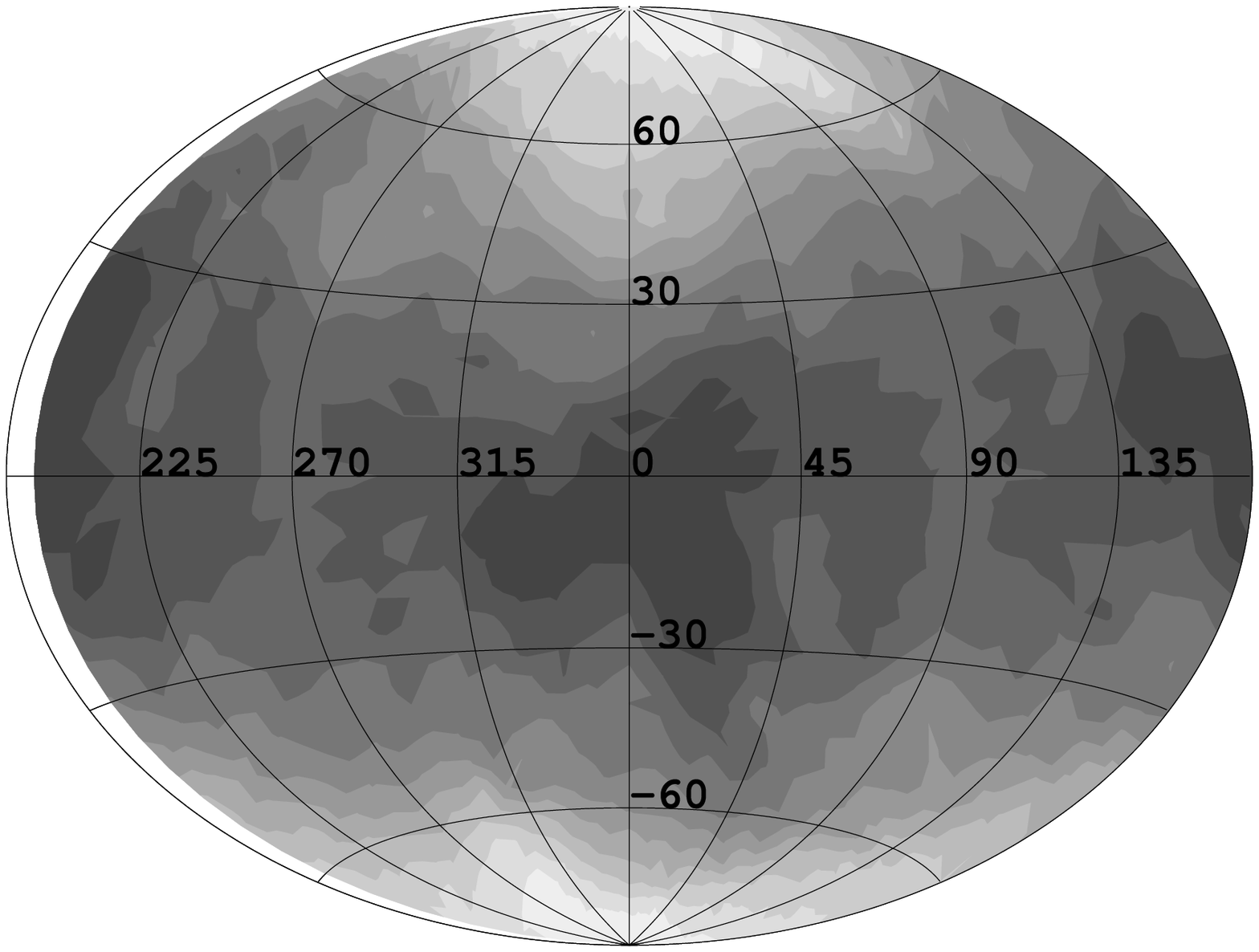}} \caption{Experimental map of the
non-uniformity of the FGC objects' normals' position angles
distribution.} \label{fgc:Amirkhanyan2_n_en}
\end{figure*}

\noindent Here $\omega$ is the angle between the plane
perpendicular to the anisotropy axis, and the axis of the radio
source. For any $\omega$ the distribution in the second coordinate
$\varphi$ is uniform within the interval from 0 to 2$\pi$. Hence
the direction $\omega=\pm\pi$/2 is the symmetry axis of the model,
which can point to any direction ($\alpha_o, \delta_o$) in the
sky. If ${\rm A} < 0$, the maximum of the distribution function
lies on the symmetry axis, and the minimum is located at
$\omega=0$ for $\varphi=0$--$2\pi$. If ${\rm A} > 0$, the maximum
and minimum replace each other. If ${\rm A}=0$, the function
$P(\varphi,\omega)=\frac{1}{\pi}$ is isotropic. The dashed line in
Fig.~3 shows the cross section of function (2). For the function
to fit the scale of Fig.~3, we normalized it
by~$\frac{1}{581.17\pi}$;

(3) for each radio source, in accordance with distribution (2) we
generate the coordinates ($\alpha_i, \delta_i$) of the direction
in the sky the axis of the source is pointed at. As a result we
obtain the map of the spatial orientation of these axes;

(4) given the coordinates of the radio sources and the
orientations of their axes, we compute the position angles of the
axes of the radio sources with respect to the Celestial Pole;

(5) we generate the catalog of the coordinates and position angles
of objects;

(6) we run the \textbf{anisoreal} code to analyze this catalog and
produce the map of non-uniformity, which can then be compared to
the given anisotropy function.

\begin{figure*}[tbp]
\setcaptionmargin{5mm} \onelinecaptionsfalse \captionstyle{normal}
\centerline{\includegraphics[width=10cm, bb=58 176 552 549,
clip]{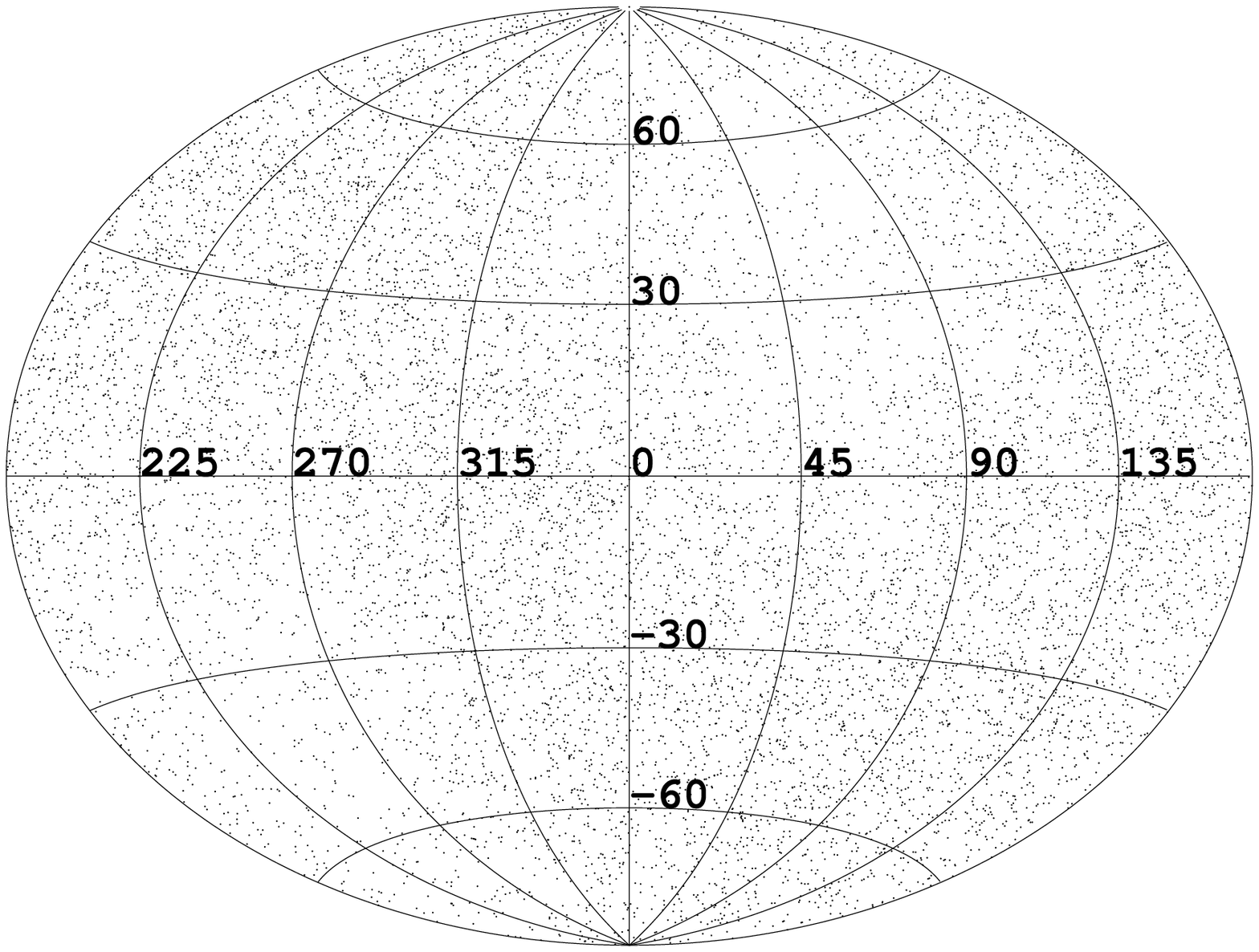}} \caption{The map of the axes
orientation of the radio sources from the simulated catalog. The
axis of anisotropy points towards \mbox{$\alpha_0=60\degr$,}
$\delta_0=45\degr$, ${\rm A}=3$. Radio sources are uniformly
distributed over the sky.} \label{osi:Amirkhanyan2_n_en}
\end{figure*}

\begin{figure*}[tbp]
\setcaptionmargin{5mm} \onelinecaptionsfalse \captionstyle{normal}
\centerline{\includegraphics[width=14cm, bb=18 288 573 719,
clip]{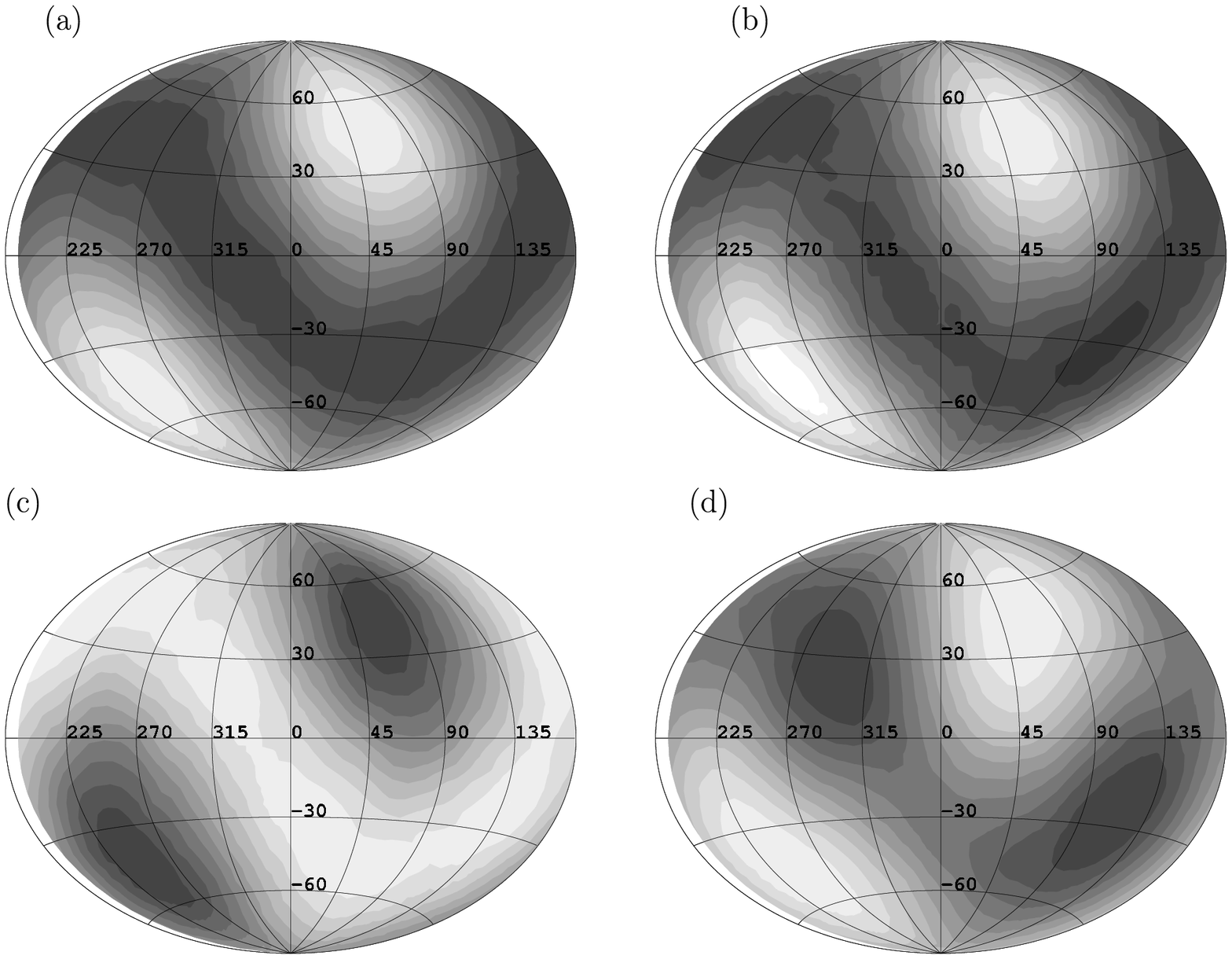}}
 \caption{Maps of the non-uniformity of the position angles distribution
 of the radio sources from the simulated catalogs. (a) The axis of anisotropy points towards
$\alpha_0=60\degr$,$\delta_0=45\degr$, ${\rm A}=3$. The radio
sources are uniformly distributed in the sky. (b) The map of the
radio sources axes orientation for the simulated catalog (Fig.~5)
folded with a low-frequency spatial filter with an aperture of
90$\degr$. (c) The anisotropy axis points towards the point
$\alpha_0=60\degr$, $\delta_0=45\degr$, ${\rm A}=-0.5$. Radio
sources are uniformly distributed over the sky. (d) The anisotropy
axis points towards the direction $\alpha_0=60\degr$,
$\delta_0=45\degr$, ${\rm A}=3$. Radio sources fill the domain of
the FIRST survey---the coordinates of the radio sources are
adopted from the catalog~\cite{9:Amirkhanyan2_n_en}. }
\label{mapmod:Amirkhanyan2_n_en}
\end{figure*}

For example, let us set the parameters of aniso\-tropy, or, more
precisely, the parameters of the spatial distribution of axes of
the radio sources (2) as follows: $\alpha_0 =60\degr, \delta_0 =
45\degr,  A=3$. We distribute 10461 radio sources uniformly over
the sky. The code generates a map of the distribution of simulated
axes (Fig.~5) and a map of non-uniformity of histograms (Fig.~6a).
As ${\rm A} > 0$, the axis of anisotropy coincides with the domain
of avoidance of the orientation of axes of the radio sources,
implying that the negative extremum of the map (Fig.~6a) must be
located near ($\alpha_0,\delta_0$). It is at this point that the
map reaches its minimum (--161.8). In the plane perpendicular to
the axis we see a dark contour corresponding to the domain of the
preferred orientation, which also corresponds to the given
anisotropy function. Such a clear pattern is not easy to discern
in Fig.~5. The explanation is very simple: the search program is
actually a strong filter suppressing low spatial frequencies,
which is applied to the data used to build the map in Fig.~6a. An
analysis of the noise of the simulated map for the isotropic case
(${\rm A}=0$) allowed us to coarsely estimate the aperture of this
filter, which is about 90$\degr$. When applied to the map of the
distribution of axes, such a filter yields a clear pattern
(Fig.~6b), the structure of which reproduces the map in Fig.~6a.
The filter improves the signal-to-noise ratio of the map, but at
the same time, as it is evident from Fig.~3, it reduces the
amplitude of anisotropy. Reversing the polarity of the anisotropy
coefficient (${\rm A}=-0.5$), results, as it follows from
\mbox{formula (2),} in a non-uniformity map with a positive
extremum in the direction of the anisotropy axis (Fig.~6c). These
numerical simulations lead us to a confident conclusion that the
\textbf{anisoreal} code adequately describes the distribution of
axes over the sky. We repeat our simulations with the same
anisotropy parameters, but with the coordinates of real
objects~\cite{9:Amirkhanyan2_n_en}, the distribution of which
reproduces the domain covered by the FIRST survey.  As a result,
we obtain the map of histogram non-uniformity (Fig.~6d), which
differs from the map in Fig.~6a by a high non-uniformity in the
domain of preferred orientation. The histogram of position angles
counted with respect to the Celestial Pole also exhibits a strong
non-uniformity and asymmetry. Repeated simulations for different
samples of random numbers and anisotropy parameters showed that
the nonuniform filling of the sky by radio sources leaves an
indelible signature in the map of the histogram non-uniformity.
The same effect explains the positive extremum in the experimental
map in Fig.~1. A trivial conclusion follows: a program built to be
used for estimating the anisotropy parameters must take into
account the distribution of radio sources in the sky. Our attempt
to construct an ideal analytical model yielded less reliable
results.

\section{PARAMETERS OF THE ANISOTROPY}

To address our task, i.e., to try to find the parameters of
anisotropy, using the most simple and natural method of
manipulating the anisotropy parameters,  $\alpha_0, \delta_0$, and
A, and minimizing the residuals between the simulated and
experimental maps of histogram non-uniformity. We construct our
model based on the actual distribution of radio sources over the
sky. We perform this operation with several samples of random
numbers in order to estimate not only the parameters, but also
their errors. As a result, we infer the following anisotropy
parameters:

$\alpha_0=180\degr\pm92\degr$, $\delta_0=84\degr\pm6\degr$, ${\rm
A}=0.48\pm0.08.$

We use these very parameters in our cross section of the space
distribution of angles in Fig.~3 (the dotted curve). The
root-mean-square deviation between the isotropic distribution and
the experimental map is equal to 29.3, and the root-mean-square
deviation between the simulated and experimental maps is 6.9. The
ratio of these standard deviations implies that the probability of
the experimental distribution could be a randomly drawn sample of
an isotropic distribution is $P<0.00004$. We used the same
programs to apply the above procedures to the catalog of edge-on
galaxies~\cite{8:Amirkhanyan2_n_en} to obtain the following parameters of
anisotropy:

$\delta_0=89\degr$, ${\rm A}=0.35$.

\noindent The right ascension $(\alpha_0)$ remains uncertain,
since the point is close to the Celestial Pole.

The probability that the spatial distribution of the normals of
galaxies is isotropic is less than 10$^{-5}$.

\section{CONCLUSIONS}

An analysis of extended radio sources position angles distribution
showed that their axes are distributed anisotropically. The axes
of radio sources avoid the direction towards the Celestial Pole
and their preferred direction is towards the equator. The ratio of
the probability densities of the orientation in these directions
is close to $0.68$. The same is true for the normals of the FGC
catalog objects distribution, however, the corresponding ratio of
probability densities is about $0.74$. We cannot propose any
mechanism that could result in such an alarming agreement of the
directions of axes in both the optical and radio experiments.
Numerical simulations show no software-related errors of such
kind. The redshift distribution of 1801 radio sources of the
catalog~\cite{9:Amirkhanyan2_n_en} leads us to cautiously conclude that the
anisotropy extends out to Z $\sim 1$ or farther. The spatial
distribution function must also undoubtedly have a finer structure
in terms of locations and types of objects. To analyze this
structure, more extensive catalogs and more sophisticated methods
are required.

\end{document}